\documentstyle[aps,prb,preprint,epsf]{revtex}
\begin{document}
\author{{\bf O. V. Danylenko}$^1$, {\bf O. V. Dolgov}$^{1,2}$,}
\address{{\bf M. L. Kuli\'c}$^3$ and {\bf V. Oudovenko}$^{4,5}$ \\ $^1$P. N. Lebedev
Physical Institute, Moscow, Russia \\ $^2$Institut f\"ur Theoretische
Physik, Universit\"at T\"ubingen, Germany \\ $^3$Centre de Physique
Th\'eorique et de Mod\'elisation, \\ Universit\'e Bordeaux I, CNRS-URA 1537
Gradignan Cedex, France \\ $^4$Max-Planck-Institut f\"ur
Festk\"orperforschung, \\ Heisenbergstr.1, 70569 Stuttgart, Germany \\ $^5$%
Joint Institute for Nuclear Research, 141980 Dubna, Russia}
\title{Normal and superconducting state in the presence of forward electron-phonon
and impurity scattering}
\date{05 September 1997}
\maketitle

\begin{abstract}
Impurities with the pronounced forward scattering (FS impurities) change
analytical properties of the quasiparticle Green's function substantially
compared to the isotropic scattering. By assuming that the superconducting
pairing is due to the forward E-P scattering (FEP pairing) it is shown that
the critical temperature of clean systems $T_{c0}$ depends linearly on the
E-P coupling constant $\lambda $ and the isotope effect $\alpha $ is small.
The FS impurities make $\alpha =1/2$ in the dirty limit and affect in the
same way the s- and d-wave FEP pairing. The FS impurities are pair-weakening
in both pairing channels. The usual isotropic impurity scattering is
pair-weakening for the s-wave and pair-breaking for the d-wave FEP pairing.
\end{abstract}

\draft

\newpage\ 

I. \underline{Introduction} - There are growing experimental evidences for
d-wave pairing in high-T$_c$ superconductors (HTS) \cite{Dwave} - seemingly
in contradiction with the standard phonon mechanism. The search for the
pairing mechanism in HTS materials has opened new directions in the theory
of superconductivity. For instance, in \cite{Pines},\cite{Scalapino} it has
been proposed the antiferromagnetic spin-fluctuation (AFS) pairing in
copper-oxides with pronounced peaks at $\vec Q=(\pm \pi ,\pm \pi )$ in the
spin-fluctuation spectral density $P_s(\vec k,\omega )$. Since the treatment
of the AFS pairing is approximate and uncontrollable, it is still unclear
which mechanism is underlying d-wave pairing in HTS systems. However, a
possibility for the phonon mechanism of superconductivity in HTS materials
has been analyzed in \cite{Shulga},\cite{Maksimov}, where the strong E-P
coupling was extracted from an analysis of optic \cite{Dolgov} and tunneling
measurements \cite{Vedeneev}. In that respect one expects that the very
sophisticated and recently developed methods of synthesis of HTS oxides \cite
{Bozovic} will give an impetus for new tunneling measurements, which might
resolve the role of the E-P interaction in the pairing mechanism.

The possibility of d-wave pairing in HTS materials due to the E-P coupling
has been put forward in a series of papers \cite{Kulic}. It has been also
studied recently in \cite{Entel}. In \cite{Kulic} it was shown that for
small hole doping $\delta $ strong Coulomb correlations renormalize the E-P
interaction giving rise to the strong forward scattering peak, while the
backward scattering is strongly suppressed. This renormalization of the E-P
coupling constant $\mid g(\vec q)\mid ^2$is described by the vertex function 
$\Gamma (\vec q)$, i.e. $\mid g_{scr}(\vec q)\mid ^2=\mid g_0(\vec q)\mid
^2\Gamma ^2(\vec q)$ , where $g_0(\vec q)$ is the bare coupling constant and 
$\Gamma ^2(\vec q)$ is strongly peaked at $\vec q=0$. Note, in what follows
we assume that the square of $\Gamma (\vec q)$ is strongly peaked at $\vec q%
=0$. This specific screening of the E-P interaction suppresses the electric
resistivity \cite{Kulic}, \cite{Grilli} and can also lead to the d-wave
superconductivity \cite{Kulic}. The physical meaning of this renormalization
is that each quasiparticle, due to the suppression of the doubly occupancy
on the same lattice - strong correlations, is surrounded by a giant
correlation hole with a characteristic size $R\simeq a/\delta $, where $a$
is the lattice constant. Later on, these results were confirmed in
calculations by slave-boson method \cite{Keller}. Some other possibilities
for the long-range E-P interaction (pronounced forward scattering), which is
due to the poor Coulomb screening in HTS oxides, have been proposed in \cite
{Abrikosov}, \cite{Weger}. In \cite{Pietronero} it was shown that in the
presence of the pronounced forward E-P interaction (FEP pairing) the vertex
(non-Migdal) corrections might increase the superconducting critical
temperature $T_c$. However, in \cite{Pietronero} only the small angle
scattering was considered, which is different from the scattering pronounced
at small $\vec q$ - which is considered here in accordance with \cite{Kulic}%
, \cite{Abrikosov}, \cite{Weger}. Recently it was argued \cite{Varelog} that
the forward E-P scattering can, in combination with the topology of the
electronic Fermi surface in HTS oxides, even produce linear $T$ dependent
resistivity down to temperatures as $10$ $K$. For further analysis it is
important fact that the nonmagnetic impurity scattering is also renormalized
by strong correlations in the similar manner as the E-P interaction \cite
{Kulic}, i.e. $u_{scr}^2(\vec q)=u_0^2(\vec q)\Gamma ^2(\vec q)$, where $u_0(%
\vec q)$ is the bare scattering potential. The pronounced forward scattering
in the pairing interaction and in the impurity scattering (FS impurities)
change the physics of the problem substantially, as it is demonstrated
below. We stress that the problem of the pronounced forward scattering is
very delicate and far from understanding, as it has been already discussed
in the framework of the phenomenological Landau's Fermi liquid \cite{Khodel}%
. As it will be demonstrated below vertex corrections become very important.
In that sense our paper might be an impetus for further studies of this
problem.

In the following analysis we assume an extreme case of the forward
electron-phonon interaction (FEP pairing) and of the forward nonmagnetic
impurity scattering (FS impurities), i.e. that $\mid g_{scr}(\vec q)\mid
^2\sim \delta (\vec q)$ and $u_{scr}^2(\vec q)\sim \delta (\vec q)$, where $%
\delta (\vec q)$ is the Dirac delta-function. We emphasize that this, rather
extreme, approximation picks up the main physics. This is also a valuable
approximation whenever the range of the effective interaction fulfills the
condition $R\gg k_F^{-1}$, i.e. $q_c(\sim 1/R)\ll k_F$. Moreover, it greatly
simplifies the structure of the Eliashberg equations by omitting integration
in $\vec k$-space. Similar approximation has been successively used in the
problem of the AF spin-fluctuation mechanism of pairing, where the four
peaks at $\vec Q=(\pm \pi ,\pm \pi )$ in the spin-fluctuation density $%
P_{sf}(\vec k,\omega )$ were replaced by four delta-functions \cite{Kostur}.
The competition of the FEP mechanism of pairing and the AFS pairing has been
studied in \cite{Licht}.

The FEP interaction and the FS impurities change physical properties in the
normal and superconducting state significantly. For instance, the FS
impurities induce substantial self-energy effects in the normal state, while
the superconducting critical temperature $T_c$ for the FEP pairing depends
linearly on the E-P coupling constant $\lambda $. We show also that the FS\
impurities affect $T_c$ for d- and s-wave FEP\ pairing equally, while they
do not affect the usual isotropic BCS pairing. In the case of the forward
scattering the vertex corrections of the E-P and of the impurity self-energy
are very important. While the vertex corrections for the FS impurities in
the normal state are considered in this paper, the E-P and the impurity
vertex corrections in the normal and superconducting state will be studied
elsewhere \cite{DDKO}.

II. \underline{Eliashberg equations for FEP pairing and FS impurities} - Let
us write the Eliashberg equations in the presence of the FEP pairing
potential $V_{ep}(\vec k,\omega )=\delta (\vec k)V_{ep}(\omega )$ and in the
presence of the FS impurities ($u_{scr}^2(\vec k)=\delta (\vec k)u^2$),
which are treated in the self-consistent Born approximation - on vertex
corrections see below. The normal and the anomalous Green's functions are
defined by $G(\vec k,\omega _n)=$ $-[i\omega _nZ(\vec k,n)+\bar \xi _n(\vec k%
)]/D(\vec k,n)$ and $F(\vec k,\omega _n)=-Z(\vec k,n)\Delta (\vec k,n)/D(%
\vec k,m)$ respectively. The wave-function renormalization $Z(\vec k%
,n)\equiv Z_n(\xi )$, the energy renormalization $\bar \xi (\vec k,n)\equiv 
\bar \xi _n(\xi )$ and the superconducting order parameter $\Delta (\vec k%
,n)\equiv \Delta _n(\xi )$ are solutions of the following equations ($\omega
_n=\pi T(2n+1)$)

$$
Z_n(\xi )=1+\frac T{\omega _n}\sum_mV_{eff}(n-m)\frac{\omega _mZ_m(\xi )}{%
D_m(\xi )}, 
$$
$$
\bar \xi _n(\xi )=\xi (\vec k)-T\sum_m\frac{V_{eff}(n-m)}{D_m(\xi )}\bar \xi
_m(\xi ), 
$$
\begin{equation}
\label{eq.1}Z_n(\xi )\Delta _n(\xi )=T\sum_m\frac{V_{eff}(n-m)Z_m(\xi
)\Delta _m(\xi )}{D_m(\xi )}. 
\end{equation}
Here, $V_{eff}(n-m)=V_{ep}(n-m)+\delta _{n,m}n_iu^2/T$ and $D_n(\xi
)=[\omega _nZ_n(\xi )]^2+\bar \xi _n^2(\xi )+[Z_n(\xi )\Delta _n(\xi )]^2$, $%
\xi (\vec k)$ is the bare quasiparticle spectrum and $n_i$ is the impurity
concentration. Let us consider first the normal state by neglecting the E-P
coupling, i.e. the self-energy is due to the FS impurities only.

III. \underline{Normal state in the presence of FS impurities only} - In
that case we put $V_{ep}(n-m)=0$ and $\Delta _n(\xi )=0$ and consider the
renormalization of the Green's function in the normal state by the FS
impurities in the self consistent Born approximation, i.e. $G^{-1}(\vec k%
,\omega _n)\equiv i\bar \omega _n(\xi )-\bar \xi _n(\xi )=G_0^{-1}(\vec k%
,\omega _n)-\Sigma ^{imp}(\vec k,\omega _n)$, where $i\bar \omega _n(\xi
)\equiv i\omega _nZ_n(\xi )$. In the self-consistent Born approximation the
self-energy is given by $\Sigma _B^{imp}(\vec k,\omega _n)=\Gamma _F^2G(\vec 
k,\omega _n)$, where $\Gamma _F=\sqrt{n_i}u$ - see $Fig.1a$. The solutions
for $\bar \xi _n$ and $\bar \omega _n$ are

$$
\bar \xi _n=\xi [\frac 12+\frac{\omega _n}{\sqrt{(\omega _n+i\xi )^2+4U}+%
\sqrt{(\omega _n-i\xi )^2+4U}}], 
$$
\begin{equation}
\label{eq.2}\bar \omega _n=\frac{\omega _n}2+\frac 14[\sqrt{(\omega _n+i\xi
)^2+4U}+\sqrt{(\omega _n-i\xi )^2+4U}]. 
\end{equation}
By using eq.(2) and for the standard isotropic spectrum one obtains $%
N(\omega _n)/N(0)=1$, i.e. the density of states in the presence of the FS
impurities is \underline{unrenormalized} - like in the case of the normal
isotropic impurity scattering (NS impurities). This might be not the case
for an anisotropic spectrum. However, the FS impurities renormalize strongly
the quasiparticle inverse life-time $\tau _B^{-1}(\omega )\equiv -Im\Sigma
_B^{imp}(\xi =0,\omega )$. By using eq.(2) and by the analytical
continuation of $\Sigma _B^{imp}(\xi =0,i\omega _n\rightarrow \omega
+i\delta )$ one obtains ($\Gamma _F=\sqrt{n_i}u$) 
\begin{equation}
\label{eq.3}\tau _B^{-1}(\omega )=\Gamma _F\Theta (2\Gamma _F-\omega )\sqrt{%
1-(\frac \omega {2\Gamma _F})^2}\text{ }. 
\end{equation}
$\Theta $ is the Heavyside function. We see that the FS impurities (in the
self-consistent Born approximation) introduce a nonanalycity of $\tau
_B^{-1}(\omega =0)$ as a function of $n_i$, i.e. $\tau _B^{-1}(\omega =0)=%
\sqrt{n_i}u$. One should compare this result with the case of the NS
impurities, where $\tau _B^{-1}(\omega =0)\sim n_iN(0)u^2$.

However, due to the absence of the integration over momenta in $\Sigma
^{imp}(\vec k,\omega _n)$ (for the FS impurities) there is no small
parameter in the theory, like $k_Fl$ for the NS impurities, where $k_F$ is
the Fermi wave vector and $l$ is the quasiparticle mean-free path.
Therefore, the vertex corrections for the FS impurities become important -
see $Fig.1b$. The summation of diagrams in $Fig.1b$ yields the self-energy $%
\Sigma _V^{imp}(\vec k,\omega _n)=\Gamma _F^2G(\vec k,\omega _n)/[1-\Gamma
_F^2G^2(\vec k,\omega _n)]$. It is seen that the vertex corrections screen
the single impurity scattering in the Born approximation. The retarded
quasiparticle Green's function $G(\xi =0,i\omega _n\rightarrow \omega
+i\delta )\equiv G_R(\omega )$ is given by 
\begin{equation}
\label{eq.4}G_R(\omega )=\frac{12^{1/6}}{\sqrt{3}\Gamma _F}[\frac \omega {%
\Phi (\omega )}+\frac 1{12^{1/3}}\frac{\Phi (\omega )}{\omega \Gamma _F^2}], 
\end{equation}
where $\Phi (\omega )\equiv \Phi (i\omega _n\rightarrow \omega +i\delta )$
and $\Phi (i\omega _n)=[9\Gamma _F^4\omega _n^2+\sqrt{(9\Gamma _F^4\omega
_n^2)^2+12\Gamma _F^6\omega _n^6}]^{1/3}$. The inverse quasiparticle
life-time $\tau _V^{-1}$ has the form 
\begin{equation}
\label{eq.5}\tau _V^{-1}(\xi =0,\omega )\approx \Gamma _F^{2/3}\omega ^{1/3},%
\text{ for }\omega \ll 3\sqrt{3}\Gamma _F/2. 
\end{equation}
It is seen in eq.(5) the nonanalycity of $\tau _V^{-1}$, i.e. $\tau
_V^{-1}\sim \omega ^{1/3}$. This result means also that $\tau _V^{-1}\ll
\tau _B^{-1}$ for $\omega \ll \Gamma _F$, i.e. the scattering by many
impurities screen the single impurity scattering. The density of states and
some quasiparticle properties in the vertex approximation will be studied
elsewhere \cite{DDKO}. Note, the above results for $\xi =0$ hold also in the
presence of the weak coupling E-P interaction.

IV. \underline{Superconductivity due to FEP and in the presence of FS
impurities} - According to the above analysis one expects that
superconductivity due to the FEP pairing might be different from the usual
BCS (or Eliashberg) pairing and that the FS and NS impurities affect this
pairing strongly. In further calculations two assumptions are made: 1. the
weak coupling E-P interaction is considered - the strong coupling studied
elsewhere \cite{DDKO}; 2. the FS impurities are studied in the
self-consistent Born approximation. It seems that the latter approximation
might be inadequate in treating superconductivity, because the vertex
corrections change the self-energy of the normal state substantially.
Moreover, the impurity vertex corrections give also anomalous vertex
corrections to the gap equation in eq.(1), which are at present
unfortunately unknown. However, the reduction of $T_c$ for the FEP pairing
by the FS impurities (in the Born approximation) is expected to be preserved
qualitatively in the superconducting state. In the latter case one expects
less reduction of $T_c$, due to the screening by the vertex corrections.

A. \underline{$T_c$ due to FEP pairing in clean systems} - We find $T_c$ in
the weak coupling limit \cite{Allen}, where $V_{ep}(n-m)\approx V_{ep}\Theta
(\Omega -\mid \omega _n\mid )\Theta (\Omega -\mid \omega _m\mid )$ and $%
\Omega $ is the phononic cut-off energy. In this limit one obtains $Z(\vec k%
,n)=1$ and the FEP pairing gives the maximum $T_c$ for $\xi =0$ on the Fermi
surface, where $\bar \xi _n(\vec k,n)=0$ . The solution of the eq.(1) in the
weak coupling limit and for $T_c\ll \Omega $ is given by $T_{c0}=\lambda
/4N(0)$, where $\lambda =N(0)V_{ep}$. Four points should be stressed. First, 
$T_{c0}$ depends on $\lambda $ linearly, what is a consequence of the
delta-function limit for $V_{ep}(\vec k,\omega )$. This result is similar to
the one obtained in \cite{Khodel} and in \cite{Kostur}, where the
delta-function limit of the AFS pairing was used\cite{Kostur}. Second, in
the AFS pairing there is an additional condition (threshold) for pairing $%
V_{sp}>2\mid \mu \mid $, where $V_{sp}$ is the spin-fluctuation constant and 
$\mu $ is the chemical potential \cite{Kostur}. Third, in \cite{DDKO} it is
shown that the expression for $T_{c0}\equiv T_{c0}(q_c=0)$ is the
zeroth-order with respect to $q_c$ - the cut-off in the momentum dependence
of the pairing potential $V_{ep}(\vec q,\omega _n)$. For $q_c\sim 2k_F$ ,
i.e. $q_cV_F\sim W\sim 1/N(0)$ one obtains the standard BCS result $%
T_{c0}^{BCS}=1.13\Omega \exp (-1/\lambda )$, while for $q_cV_F\ll \Omega $
(the FEP pairing) the correction to $T_{c0}$ is given by $T_c\simeq
T_{c0}\left( 1-7\zeta (3)q_cV_F/4\pi ^2T_{c0}\right) $. Finite $q_c$ lowers $%
T_{c0}$ of the FEP pairing. Fourth, in the weak coupling limit there is no
isotope effect ($-d\ln T_{c0}/d\ln M=$ $0$ for $\Omega \rightarrow \infty $%
), although the pairing is due to the E-P interaction. This is a consequence
of the FEP mechanism of pairing. The strong coupling effects introduce the
mass ($M$) dependence of $T_c$. For example, by assuming the single mode
(Einstein) phononic spectrum with the frequency $\Omega $ then the effective
interaction in eq.(1) has the form $V_{eff}(\omega _n)=V_{ep}\Omega
^2/(\Omega ^2+\omega _n^2)$. By assuming that $[\lambda N^{-1}(0)/2\Omega
]<1 $ one obtains a correction for $Z_n(\xi =0)$ (note $\bar \xi _n(\xi =0)=0
$) 
\begin{equation}
\label{eq.6}Z_n(0)\simeq 1+\frac{\lambda N^{-1}(0)}{2\Omega }\frac{\Omega ^2%
}{\Omega ^2+\omega _n^2}. 
\end{equation}
Because $Z>1$ it lowers the critical temperature $T_{c0}^{(1)}=T_{c0}/[1+%
\lambda /2\Omega N(0)]^2$, i.e. $T_{c0}^{(1)}<T_{c0}$. Since $\Omega \sim
M^{-1/2}$ one obtains the finite isotope effect $\alpha \equiv -(d\ln
T_{c0}^{(1)}/d\ln M)\approx 2T_{c0}/\Omega $. In obtaining $T_{c0}$ it is
assumed $T_{c0}\ll \Omega $, which gives $\alpha \ll 1/2$. The problem of
the isotope effect is much more complicated than it is treated here.
However, this simplified analysis gives an impetus for future study.

B. \underline{$T_c$ due to FEP in the presence of FS impurities} - The FS
impurities affect $T_c$, which is in the weak coupling limit given by 
\begin{equation}
\label{eq.7}1=V_{ep}T_c\sum_{\omega _n=-\Omega }^\Omega \frac 1{\omega
_n^2Z_n(\xi =0)}, 
\end{equation}
where $Z_n(\xi =0)=(1+\sqrt{1+4\Gamma _F^2/\omega _n^2})/2$ and $\bar \xi
_n(\xi =0)=0$. Note, eq. (7) holds for all kind of FEP pairings (s-,d-,
etc.). Some limiting cases are considered:

(a) \underline{$\Gamma _F\ll \pi T_c$} : One obtains $T_c=T_{c0}[1-4\Gamma
_F/49T_{c0}]$. Note, the singular slope $-dT_c/dn_i\sim n_i^{-1/2}$ ; (b) 
\underline{$\Gamma _F\gg \pi T_c$} : If $\Gamma _F\gg \Omega /2$ is
fulfilled one obtains $T_c\approx (\pi /2\gamma )\Omega \exp (-\pi \Gamma
_F/V_{ep})$, $\gamma \approx 1.78$.

We point out three results. First, the slope $-dT_c/dn_i$ depends on $n_i$,
contrary to the case of usual d-wave superconductors with the NS impurities.
This is because $\Gamma _F=\sqrt{n_i}u$ for the FS impurities instead of $%
\Gamma =n_iN(0)u^2$ for the NS impurities. Second, the FS impurities are 
\underline{pair weakening} for the FEP pairing due to the exponential
fall-off of $T_c$ with the increase of $\Gamma _F.$ This is contrary to the
effect of the NS impurities on usual d-wave pairing, where $T_c=0$ for $%
\Gamma _{cr}\approx 0.8T_{c0}$ - the pair breaking effect. Third, the FS
impurities give rise to the large isotope effect $\alpha =1/2$ in the dirty
limit $\Gamma _F\gg T_c$, although $T_{co}$ is mass independent.

C. \underline{$T_c$ due to FEP in the presence of NS impurities } - In this
case one should make the difference between s-wave and d-wave FEP pairings.
For the NS impurities one has $\bar \xi _n(\xi )=0$ and $Z(\vec k,n)$ is
isotropic, i.e. one has $Z_n=1+\Gamma /\mid \omega _n\mid $, where $\Gamma
=\pi N(0)u^2$.

(1) \underline{d-wave FEP pairing} - $T_c$ in the weak coupling limit is
given by

\begin{equation}
\label{eq.8}1=V_{ep}T_c\sum_{\omega _n=-\Omega }^\Omega \frac 1{\omega
_n^2Z_n^2}. 
\end{equation}

(a) \underline{$\Gamma \ll \pi T_c$} : One obtains $T_c=T_{c0}[1-2\Gamma
/\pi T_{c0}]$. Note, in the case of the d-wave FEP\ pairing the slope $%
-dT_c/d(\Gamma )=2/\pi $ is smaller than the slope of the usual d-wave
pairing in the presence of the NS impurities, where $-dT_c/d(\Gamma )=\pi /4$%
. As a consequence the d-wave FEP\ pairing is \underline{more robust} in the
presence of the NS\ impurities than the usual d-wave pairing; (b) \underline{%
$\Gamma \gg \pi T_c$} : In this limit one obtains from eq.(8) that $T_c=0$
for $\Gamma _{cr}\simeq (4/\pi )T_{c0}$. Note, in the case of the usual
d-wave pairing $\Gamma _{cr}\simeq 0.8T_{c0}$, which confirms our statement
on the robustness of the d-wave FEP pairing, compared with the usual d-wave
pairing.

(2) \underline{s-wave FEP pairing} - $T_c$ in the weak coupling limit has
the form

\begin{equation}
\label{eq.9}1=V_{ep}T_c\sum_{\omega _n=-\Omega }^\Omega \frac 1{\omega
_n^2Z_n}. 
\end{equation}
Note, the denominator in eq.(9) is proportional to $Z_n$ while for the
d-wave FEP pairing to $Z_n^2$ - see eq.(8). Since $Z_n>1$ it means that in
the presence of the NS impurities the s-wave FEP pairing is \underline{more
robust} than the d-wave FEP pairing. By solving eq.(9) one obtains 
\begin{equation}
\label{eq.10}\frac{T_c}{T_{c0}}=\frac 4{\pi ^2\rho }\psi (\frac 12+\frac \rho
2)-\psi (\frac 12), 
\end{equation}
where $\psi (x)$ is the di-gamma function and $\rho =\Gamma /\pi T_c$.

(a) \underline{$\Gamma \ll \pi T_c$} : One obtains $T_c=T_{c0}[1-7\zeta
(3)\Gamma /\pi ^3T_{c0}]$. Note that $-dT_c/d(\Gamma )\mid _s\ll
-dT_c/d(\Gamma )\mid _d=2/\pi $, i.e. in the presence of the NS impurities
the s-wave FEP\ pairing is more robust than the d-wave FEP\ pairing; (b) 
\underline{$\Gamma \gg \pi T_c$} : In this case $T_c$ goes to zero
asymptotically, i.e. $T_c\approx (\Gamma /2\pi )\exp (-\pi \Gamma /4T_{c0})$%
. This means that the NS impurities are \underline{pair-weakening} for the
s-wave FEP\ pairing.

D. \underline{$T_c$ for isotropic E-P pairing in the presence of FS
impurities} - For completeness we consider the effect of the FS impurities
on $T_c$ of the usual BCS s-wave pairing. Based on eq.(1) one obtains $T_c$
in the presence of the FS impurities $1=V_{ep}T_c\sum_n\Theta (\Omega
-|\omega _n|)Q(\omega _n,\Gamma _F)$. One can show that $Q(\omega
_{n^{\prime }},U)=\pi /\mid \omega _n\mid $, which is the simple BCS formula
in the absence of impurities. This means that the isotropic s-wave pairing
is unaffected by the FS impurities - the Anderson theorem holds.

V. \underline{Conclusions}{\bf \ - }In summary, it is shown here that: $(a)$
impurities with the pronounced forward scattering (FS impurities) change the
analytical properties of the quasiparticle Green's function substantially
and the vertex corrections, due to the interference scattering by other than
one impurity, \underline{screen} the single impurity Born scattering. $(b)$
by assuming that the pairing is due to the forward E-P scattering the
critical temperature of clean systems $T_{c0}$ depends linearly on the E-P
coupling constant $\lambda $ in the Migdal approximation; $(c)$ the isotope
effect is small in the weak coupling limit, i.e. $\alpha \ll 1$ for $%
T_{c0}\ll \Omega $, while the FS impurities can push it to $\alpha =1/2$ in
the dirty limit ($\Gamma _F\gg T_c$); $(d)$ the FS impurities affect in the
same way the s- and d-wave FEP pairing. Moreover, the FS impurities are
pair-weakening in both pairing channels; $(e)$ the s- and d-wave FEP pairing
are differently affected by the isotropic impurity scattering (the NS
impurities). The NS impurities are pair-weakening for the s-wave FEP
pairing, while they are pair-breaking for the d-wave FEP pairing; $(f)$ the
FS impurities do not affect the usual BCS s-wave pairing.

Finally, we point out that the (non-Migdal) vertex corrections, due to the
E-P interaction, are important for the FEP pairing, because due to the
absence of momentum integration in the self-energy there is no small
parameter in the theory, like $\lambda \omega _D/E_F$ in the Migdal theory.
In that sense the case of the FS\ impurities is very instructive, because it
tells us that vertex corrections change the self-energy substantially. The
vertex corrections in the presence of the FEP scattering can increase $T_c$
as it has been asserted in \cite{Pietronero}. However, as we have shown
above the critical temperature $T_c$ in the Migdal approximation is
nonexponential function of the E-P coupling constant $\lambda $ contrary to
the exponential dependence obtained in \cite{Pietronero}. The latter is due
to the assumed small angle scattering in \cite{Pietronero}, which is very
different from the pronounced small $\vec q$ scattering studied here. The
effect of the vertex corrections on $T_c$ for the FEP pairing will be
studied elsewhere \cite{DDKO}.

{\bf Acknowledgments} - We would like to thank Nils Schopohl and E. G.
Maksimov for discussions and support. O. V. D. and O. V. D thank RFBR
(projects 96-02-16661a and 96-15-96616) and Ministry of Russian Science
(project ''HTSC-64'') for support. M. L. K. acknowledges University of
Bordeaux for kind hospitality and thanks A. I. Buzdin, L. Hedin, M. Mehring
and Y. Leroyer for support. This work was started during our stay at ISI,
Torino, Italy, October 1996 and we acknowledge the ISI support.

\newpage
FIGURE CAPTIONS

$Fig.1$

$(a)$ the self-energy $\Sigma _B^{imp}(\vec k,\omega _n)=\Gamma _F^2G(\vec k%
,\omega _n)$ in the presence of the FS impurities in the Born approximation;

$(b)$ the self-energy $\Sigma _V^{imp}(\vec k,\omega _n)=\Gamma _F^2G(\vec k%
,\omega _n)/[1-\Gamma _F^2G^2(\vec k,\omega _n)]$ in the presence of the FS
IMPurities with vertex corrections.

\newpage

\noindent
\mbox{}\\[2cm]
\parbox[h]{1.5cm} {
$$ \Sigma_{B}(\bf k,\omega_n)\quad= $$\\[18cm] }\hspace*{2.0cm} 
\parbox[t]{4.5cm} {
\vspace*{-10.8cm}
\epsfysize=8.8in
\epsffile{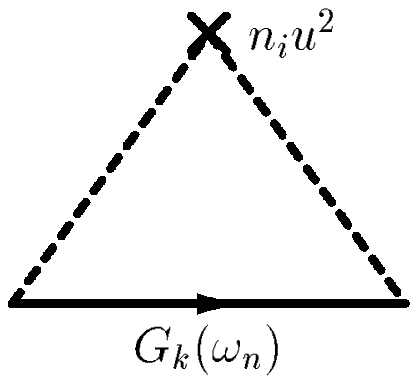}
\mbox{}\\[-22.0cm]
\hspace*{5cm}
\mbox{\Large\it a) }
}\\[-16cm]
\vspace*{-13.00cm} \noindent
\parbox[h]{1.5cm} {
$$ \Sigma_{V}(\bf k,\omega_n)\quad= $$\\[20cm] }\hspace*{2.6cm}\\[-10cm]
\vspace*{-18.6cm} \hspace*{3.6cm} 
\parbox[h]{8.5cm} {
\epsfysize=8.8in
\epsffile{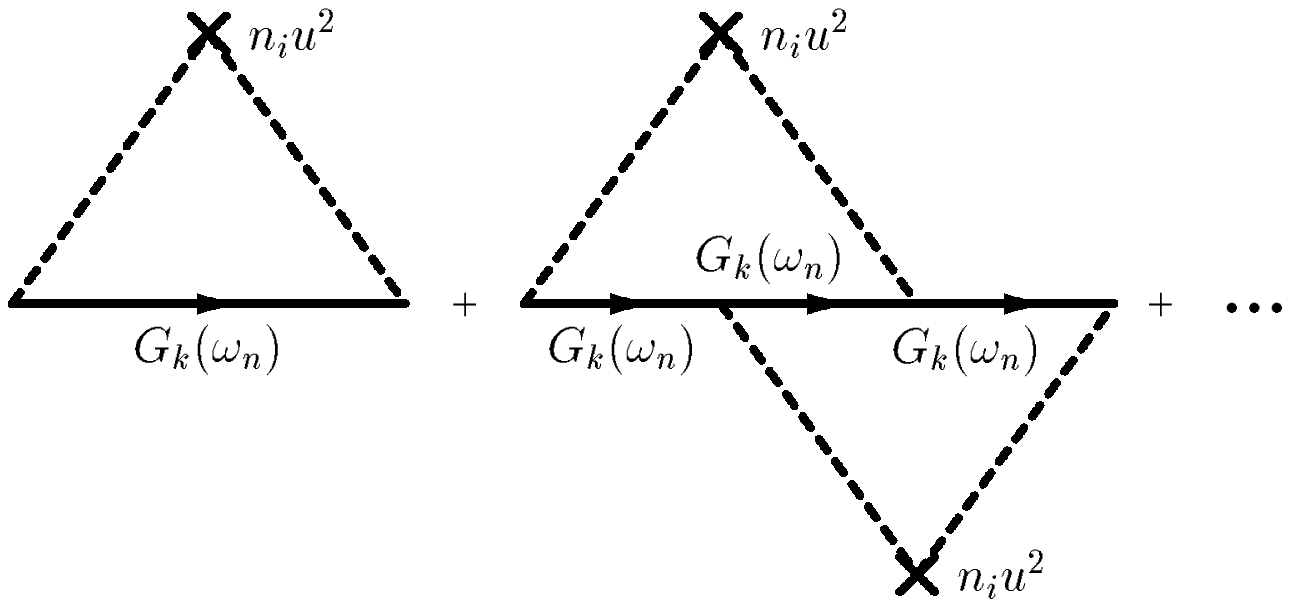}
\mbox{}\\[-20.5cm]
\hspace*{11.6cm}
\mbox{\Large\it b) }\\
}\\

\end{document}